\documentclass[12pt]{article}
\usepackage{graphicx,a4} 
\usepackage{amssymb}
\usepackage{latexsym}
\begin{document}

\newcommand{\preprintno}[1]
{\vspace{-2cm}{\normalsize\begin{flushright}#1\end{flushright}}\vspace{1cm}}

\title{\preprintno{{\bf ULB-TH/02-35}}
Varying $\alpha$ and black hole entropy}
\author{{Malcolm Fairbairn\thanks{E-mail: mfairbai@ulb.ac.be}}  and
{Michel H.G. Tytgat\thanks{E-mail: mtytgat@ulb.ac.be}}\\
{\em Service de Physique Th\'eorique, CP225}\\
{\em Universit\'e Libre de Bruxelles}\\
{\em B-1050 Brussels, Belgium}}

\date{\today}

\maketitle
\begin{abstract}
Recently it has been suggested that an increase in the fine structure constant $\alpha$ with time would decrease the entropy of a Reissner-Nordstrom black hole, thereby violating the second law of thermodynamics. 
In this note we point out that, at least for a class of charged dilaton black hole related to string theory, the entropy does not change under adiabatic variations of $\alpha$ and is expected  to increase for non-adiabatic changes.
\end{abstract}

Motivated by recent claims \cite{webb} that the fine structure constant could have been increasing with time,  Davies {\em et al} \cite{Davies:un} made the observation that such an increase should lead to a decrease in the entropy of a Reissner-Nordstrom black hole, thereby apparently violating the second law of thermodynamics.\footnote{They actually argued that  $\alpha$ could increase either because $e$, the electric charge, is increasing or because $c$, the speed of light, is decreasing and that the change in the entropy of charged black hole might allow one to discriminate between these seemingly inequivalent situations. We choose not to contribute to that aspect of the debate here. See \cite{duff} . } (For related work on this topic see \cite{others}).  
In this letter we aim to show that the entropy is not changed by slow {\em i.e.} adiabatic variations of $\alpha$ in theories where the gauge couplings are set by the expectation values of scalar fields, such as string theories.  We then go on to show why we expect the entropy of black holes in such theories to always increase for non-adiabatic changes in the scalar fields.

Let us first review the argument of Davies {\em et al}. The entropy of a Reissner-Nordstrom black hole of ADM mass $M$ and electric charged $Q$, in units in which $c=\hbar=G=1$ is given by 
\begin{equation}
\label{SRN}
S \equiv {A\over 4} =  \pi \left ( M + \sqrt{M^2 - Q^2}\right)^2
\end{equation}
where $A$ is the area of the black hole horizon.  Since $Q$ is quantized in units of the electric charge $e$, an increase in $\alpha=e^2$ appears to lead to a decrease in the entropy of the black hole (all other things being kept constant). 

This argument is however incomplete. To decide whether the entropy is changing, one must specify the process through which it could do so. For instance, when we do work on a gas by very slowly compressing it, we have to specify if heat is exchanged with the surrounding medium. If not, the temperature and internal energy of the gas rises but the entropy stays constant. Similarly an adiabatic variation in $\alpha$  should change the mass of a charged black hole but not, by definition,  its entropy.\footnote{See also the recent \cite{Das:2002vh} for a similar argument.}

Accordingly, the first law of black hole thermodynamics should be extented to take into account the work done by varying $\alpha$, {\em i.e.} 
$$
d M = \ldots +  {\partial M\over\partial \alpha} d \alpha
$$ 
However, if $\alpha$ is dynamical, consistency requires that we work within the framework of some Brans-Dicke type of theory. This problem has actually been solved quite some time ago, at least for a certain class of dilaton black hole related to string theory. We believe that much of what we are saying here has been said before, see in particular \cite{Gibbons:1996af}.
 
We will focus on electrically charged, dilaton black hole solutions of the following string inspired 4-dimensional effective action \cite{Gibbons:ih,gg,horowitz}  
\begin{equation}
\label{action}
I = {1\over 16 \pi}\int d^4x \sqrt{-g} \left( {R} - 2 \partial_\mu \phi \partial^\mu \phi - e^{- 2 \phi} F_{\mu\nu}F^{\mu\nu}\right)
\end{equation}
however, due to the qualitative arguments presented above, we predict a similar result for any charged black hole where the value of $\alpha$ is set by a dynamical scalar field.  We put no potential for the dilaton field $\phi$ and the asymptotic expectation value of the dilaton field $\phi_0$ is a modulus field, with $\alpha = \exp(2 \phi_0)$. A convenient parameterization of the black hole metric is \cite{Gibbons:ih}
$$
ds^2 = - e^{2 U(\tau)} dt^2 + e^{-2 U(\tau)}\left[ {c^4 \over \sinh^4{c \tau}} d\tau^2 + {c^2 \over \sinh^2 c\tau} d\Omega\right]
$$
where $\tau \rightarrow 0^-$ corresponds to asymptotic infinity and $\tau \rightarrow - \infty$ corresponds to the black hole horizon. The usefulness of this parametrization lies in the fact that $c = 2 S T_H$ where $S=A/4$ is the black hole entropy and $T_H$ is the Hawking temperature. Non-rotating electrically charged dilaton black holes can be characterized by their ADM mass $M$, electric charge $Q$ and by $\phi_0$. We give the solution of the field equations for reference: 
$$
e^{- 2 U} = \left[\frac{\sinh c (\tau_b - \tau)}{\sinh(c \tau_b)}\right] e^{-c \tau} \qquad ; \qquad e^{- 2 \phi} = \left[{\sinh c(\tau_b -\tau)\over \sinh c\tau_b}\right] e^{- 2 \phi_0 + c\tau} 
$$
where $\tau_b$ is an integration constant which satisfies
$$
e^{\phi_0} \sqrt{2} Q \sinh c\tau_b = c
$$
>From these solutions one may extract the ADM mass,
$$
M \equiv \left.{d U\over d\tau}\right\vert_{\tau =0} = {1\over 2} c \left( 1+ \coth c\tau_b\right)
$$
and the dilaton charge $\Sigma$,
$$
\Sigma \equiv -\left.{d \phi\over d\tau}\right\vert_{\tau=0} = {1\over 2} c \left (1 - \coth c \tau_b\right) \equiv - {Q^2 \over 2 M} e^{2 \phi_0}.
$$
A nice relation between $M$, $Q$ and $\Sigma$ is \cite{Gibbons:1996af}
$$
M^2 + \Sigma^2 - Q^2 e^{2 \phi_0} = c^2
$$
which essentially states that for extremal black hole,  $c=0$, the attractive gravitational and dilaton channels are compensated by the repulsive electric one. Finally, the black hole entropy and temperature read
\begin{equation}
\label{Sdilaton}
S = 4 \pi M^2 \left( 1 + {\Sigma\over M}\right) = 4 \pi M^2 \left( 1 - {Q^2 e^{2\phi_0}\over 2 M^2}\right) 
\end{equation}
and
$$
T_H = {1\over 8 \pi M}
$$
Comparing (\ref{SRN}) with (\ref{Sdilaton}), we first see that the expression of the entropy is quite different depending on whether the dilaton is massless or if it is frozen, as in the Reissner-Nordstrom case.  Despite this, repeating the argument of Davies {\em et al}, we would naively conclude that if $\alpha \equiv \exp(2 \phi_0)$ increases, the entropy of the electrically charged dilaton black hole decreases, as in the Reissner-Nordstrom case.  However this neglects the fact that the ADM mass is itself affected by a change in the asymptotic value of $\phi_0$. It can be proved using standard techniques \cite{Gibbons:1996af,Heusler:1993cj} that
\begin{equation}
\label{workofalpha}
\left.{\partial M\over \partial \phi_0}\right\vert_{S,Q} = - \Sigma 
\end{equation}
so that the derivative of equation (\ref{Sdilaton}) with respect to $\phi_0$ is zero and the correct form of the first law for non-rotating charged dilaton black holes is \cite{Gibbons:1996af}
$$
d M = T_H dS + \Phi dQ - \Sigma d\phi_0
$$
This essentially completes our claim that for adiabatic variations of $\phi_0$ ({\em i.e.} of $\alpha$), the black hole entropy does not change. 

For the case of a neutral black hole $Q=0$ and therefore $\Sigma=0$ and one recovers the Schwarschild solution with a constant dilaton field $\phi(\tau)=\phi_0$ and $U = c\tau$. If we reinstate Newton's constant, the black hole entropy is then simply given by 
$$
S = 4 \pi G M^2
$$
and is independent of $\phi_0$.  It is illuminating to compare this result obtained from the Einstein frame action (\ref{action}) to the expression one would find working in the Jordan or string frame. These two frames are related by the conformal transformation
$$
g_{\mu\nu}^S = e^{2 \phi_0}  g_{\mu\nu}^E
$$
In the Einstein frame, the gravitational coupling constant is fixed ({\em e.g.} we set it to $G=1$), whereas in the Jordan frame it depends on $\phi_0$. The two are related by
$$
G_S = e^{2 \phi_0} G
$$
However, in the string frame, the mass of the black hole also depends on $\phi$ since the scale with respect to which masses are defined, {\em i.e.} time,  is changing. In particular, 
$$
{M_S\over M} = {d t\over dt_S} = e^{-\phi_0}
$$
Note also that this implies that in the string frame the horizon radius and area also change as $\phi_0$ varies. Nevertheless, in either frame,
$$
S = 4 \pi G_S M_S^2 = 4 \pi G M^2
$$
and $S$ is independent of $\phi_0$ \cite{polchinski}. In this simple setting we can use a result derived by  Jacobson \cite{Jacobson:1999vr} to consider the non-adiabatic case. If it is not infinitesimally small, any time variation in $\phi_0$ represents a pulse of dilaton radiation which enters the black hole.  The rate of change of the black hole mass in the Einstein frame is given by the flux of energy across the horizon 
$$
{d M\over dt} \propto \left({d\phi_0\over dt}\right)^2 > 0
$$
which is always positive as expected. That is the entropy of the black hole increases for non-adiabatic, in other words not infinitesimally slow, variations in the dilaton field no matter whether $\phi_0$ is increasing or decreasing.  Presumably this result can be extended to the charged dilaton black hole but we have not tried to do so.

\section*{Acknowledgements}
We are grateful for conversations with Tom Dent, Jean-Marie Fr\`{e}re and Philippe Spindel.  This work is supported by the IUAP program of the Belgian Federal Government. MF is funded by an IISN grant and MT is funded by the FNRS.

\end{document}